# A Framework for Understanding AI-Induced Field Change: How AI Technologies are Legitimized and Institutionalized

Benjamin Cedric Larsen
Department of Economics
Government & Business
Copenhagen Business School
Copenhagen, Denmark
bcl.egb@cbs.dk

## ABSTRACT

Artificial intelligence (AI) systems operate in increasingly diverse areas, from healthcare to facial recognition, the stock market, autonomous vehicles, and so on. While the underlying digital infrastructure of AI systems is developing rapidly, each area of implementation is subject to different degrees and processes of legitimization. By combining elements from institutional theory and information systems-theory, this paper presents a conceptual framework to analyze and understand AI-induced field-change. The introduction of novel AI-agents into new or existing fields creates a dynamic in which algorithms (re)shape organizations and institutions while existing institutional infrastructures determine the scope and speed at which organizational change is allowed to occur. Where institutional infrastructure and governance arrangements, such as standards, rules, and regulations, still are unelaborate, the field can move fast but is also more likely to be contested. The institutional infrastructure surrounding AI-induced fields is generally little elaborated, which could be an obstacle to the broader institutionalization of AI-systems going forward.

## 1 Introduction

In recent years, the scope of information technology that complements or augments human actions has expanded rapidly. The logics embedded in AI-systems already operate in diverse areas, such as the stock market [1], mortgage underwriting [2], autonomous vehicles [3], medical services [4], the judicial system [5], and a range of other fields. The action-potentials inherent in most AI systems imply a shift in agency, moving from human actors to AI agents, which in turn has a significant impact on shaping new practices (e.g., across healthcare, agriculture, autonomous vehicles, etc.), and thereby new forms of organization.

Novel AI systems and agents are embedded in existing digital infrastructures and operate within an institutional framework that enables or constrains various activities [6]. The socio-economic embeddedness of AI systems means that some AI agents may affect and alter existing social practices and ways of organization in swift and transforming ways, while implementation may be subject to varying degrees of legitimacy, depending on the field and area of implementation. Digital infrastructures, however, tend to emerge more rapidly than institutional infrastructures (e.g., laws and regulations), which is commonly referred to as the pacing problem [7]. This may create extensive issues if negative externalities are associated with fast-moving technological implementation that is at odds with existing structures or norms for certain actors or groups of a population [8, 9]. Tensions also arise as human actions increasingly have become subject to 'informatization' where behavior is tracked, sometimes unknowingly, through the collection of new data points [10, 11, 12]. Data is derived from social networks and online interactions, facial recognition technologies, driving behavior, apps recording location data, and so on. The wide range of AI implementations and some of the associated tensions captured by the pacing problem, guides and motivates the research question of this paper, which seeks to understand how AI-induced fields are subject to varying degrees of legitimacy as well as processes of institutionalization.

Views from institutional- and information systems (IS) theory are combined in order to conceptualize how AI fields operate at the meso-level in terms of gaining legitimacy, that is, how AI diffusion is adopted and accepted, or rejected, under varying socio-economic conditions.

Elements from information systems theory elaborate on the notion of digital infrastructure [13, 14, 15], which signifies a range of interconnected technologies (e.g., Internet, Platforms, IoT) that contribute to realize the action potentials of novel AI agents and associated processes of information collection.

Institutional theory introduces the concept of fields, which is applied in order to denote distinct areas of AI implementation and organization by a diverse range of actors. Elements from institutional theory, i.e., institutional work [16, 17], logics [18], and infrastructure [19], are applied in order to conceptualize how processes of AI-induced digitization affects the evolution and governance of organizations [20]. Theory surrounding institutional work is applied in order to understand how actors accomplish the social construction of logics (i.e., rules, scripts, schemas, and cultural accounts), which signifies where human actors and AI agents may challenge existing organizational or institutional practices and boundaries, which may result in difficulties associated with legitimization. Adding the institutional perspective is about how "digitally-enabled institutional arrangements emerge and diffuse both through fields and organizations" [19: 53]. The primary focus of the paper is placed on the interplay between existing and new and emerging institutional arrangements, as well as the role of AI in altering ways of organization.

In combining views from institutional- and information systems (IS) theory, the paper proposes a novel conceptual framework for analyzing and understanding AI-induced field change. The framework builds on Zietsma et al.'s. [22] concept of pathways of change, which outlines how a field is likely to move between states from emerging/aligning to fragmented, contested, and established, depending on the coherency in logics and elaboration of institutional infrastructure. The proposed framework adds the notion of digital infrastructure elaborated through the constructs of technological maturity, data, and AI autonomy, which enables an assessment of the impact of AI-systems on existing forms of institutional infrastructure. Where digital and institutional infrastructure is well-elaborated in terms of organizational practices, rules, and processes, the field could be considered established. If a field is emerging or aligning, on the other hand, its digital and institutional infrastructure will be nascent and unelaborate. The developed framework is illustrated through application to the field of facial recognition technologies in the United States.

The paper contributes by elaborating on existing information systems theory through adding the institutional perspective to understand the dispersion of AI technologies. Clarity is gained in terms of assessing how AI technologies move within and between fields, which is interpreted through a technology's elaboration of institutional and digital infrastructure, which in combination informs a technology's perceived degree of legitimacy.

The paper is structured as follows. Section 2 elaborates on institutional theory and the characteristics of digital infrastructure. Section 3 presents a framework for understanding AI-induced field change. Section 4 applies the framework through illustration. Section 5 deliberates on pathways of change referring to how AI-fields become institutionalized, and section 6 discusses obstacles to legitimacy as well as paths forward in terms of governance. Section 7 concludes.

## 2  Institutional Theory and AI Agents

In organization theory, the idea of institutional infrastructure reflects understandings of the embeddedness of organizations within fields and the structuration of fields that occurs through interactions and institutional activity amongst actors [23]. Over the last few decades, organizational fields have become more dynamic, and boundaries between fields have become more porous due to the introduction of new digital infrastructures, such as the Internet [18: 336].

Early institutional theory developed the notion that organizations come to resemble each other due to socio-cultural pressures, which provide a source of legitimacy [24]. A central process is that of isomorphism, demonstrating that organizations are likely to converge through normative, mimetic, and coercive pressures [25]. Mimetic isomorphism holds that organizational legitimacy is achieved through copying other organizations as well as their technologies and practices. Coercive legitimacy refers to societal legitimacy, which often is achieved through legislation, whereas normative legitimacy can be viewed as the appropriate professional standards as well as social acceptance of new technologies. Socio-cultural beliefs and practices thus play an important role in the adoption of new technologies and innovations, as well as contingent processes of legitimization [21].

Competing institutions may lie within individual populations that inhabit a field, while fields may be contested by multiple, and often competing, institutional logics [15, 24, 25, 26, 27]. Institutional logics describe the "socially constructed, historical patterns of material practices, assumptions, values, beliefs, and rules" of a field [28: 804]. The institutional logics perspective deals with the interrelationships among individuals, institutions, and organizations, i.e., the actors of a field.

Institutional work, on the other hand, emphasizes a conceptual shift towards individuals and organization's actions as "dependent on cognitive (rather than affective) processes and structures and thus suggests an approach… that focuses on understanding how actors accomplish the social construction of rules, scripts, schemas, and cultural accounts" [14: 218].

When the two approaches are held together, i.e., logics and interrelationships, and structures and practices, these can be expressed as the institutional infrastructure of a field. Institutional infrastructure is established through adjacent activities such as certifying, assuring, and reporting against principles, codes, and standards, as well as through the formation of new associations and networks among organizations, including official rules and regulations [31]. Institutional infrastructure can be clarified in terms of its degree of elaboration (high, low), as well as coherency in logics (unitary, competing) [19].

Novel AI agents operating in varying systems also embody distinct logics and cognitive functions [32]. While these functions are defined by human actors (e.g., engineers in a company), AI-agents remain subject to different degrees of autonomy, i.e., they are to some extent able to act independently based on intrinsic flows of information. This implies that AI agents have the autonomy to act on (e.g., judicial evidence, road conditions, etc.), as well as interact with (e.g., speech recognition, chatbots) their environments. This new form of artificial agency confounds the paradox of embedded agency, i.e., how actors are able to change institutions when their actions are conditioned by those same institutions [33], by the implication of an AI's ability to shape human behavior as well as ways of organization – sometimes simultaneously. In other words, algorithms can affect how we conceptualize the world while modifying socio-political forms of organization [34].

Algorithms can be seen as non-human agents endowed with the ability to evaluate, rank, and reward or punish individuals' actions and positions based on pre-programmed instructions that shape social relationships [33, 34]. Algorithms, however, are oftentimes compressed and hidden, and we do not encounter them in the same way that we encounter traditional rules [35, 36]. The increasing reliance on algorithms as instruments for the regulation of social relationships, coupled with the obscurity of algorithmic evaluation systems, is evidence of new yet subtle ways of exercising power, which alters existing power-dependencies, e.g., through surveillance, online interaction, and so on [33, 37]. Algorithms are therefore implicated in the constitution and reproduction of power asymmetries that regulate individuals' behaviors and ensure their compliance with predefined standards, which in turn can affect human agency [35]. It is difficult, however, to identify ex-ante what the socio-economic effects of scaling an AI-system will be [38, 39], which warrants that extensive experimentation through application may be necessary before AI-based technological diffusion and legitimization is likely to take place.

Institutional logics and institutional work provide a foundation to understand the rationalities and practices of actors that implement novel AI-agents, as well as the AI-agents' systemic impact on their surroundings through their socio-economic embeddedness. An analysis of AI-agents predicated on institutional work and logics can be placed either at the micro-level, seeking to understand the impact of individual AI-agents on specific socio-economic practices, or at the meso-level, seeking to understand how actors influence the legitimacy of AI applications in a field. That is, how AI diffusion is adopted and accepted, or rejected, under varying socio-economic and technological conditions.

## 2.1 Digital Infrastructure

Digital infrastructure is made from a multitude of digital building blocks and is defined as the computing and network resources that allow multiple stakeholders to orchestrate their service and content needs [14]. Digital infrastructures are distinct from traditional infrastructures because of their ability to collect, store, and make digital data available across a large number of systems and devices simultaneously [14]. Examples of digital infrastructures include the Internet [40, 41]; data centers; open standards, e.g., IEEE 802.11 (Wi-Fi), as well as consumer devices such as smartphones.

Henfridsson et al. [38: 90] refer to "digital resources" as entities that serve as building blocks in the creation and capture of value from information. While AI technologies are assembled as digital building blocks, a distinction needs to be made between traditional software systems (i.e., ERP, CRM, WordPress, etc.) and novel AI-systems (computer vision, machine learning, etc.). This distinction is important as a new kind of embedded agency is inherent in most AI systems, which render these as "organizers," "predictors," or "controllers" of data flows that are captured by digital infrastructures [44].

Most digital building blocks are made accessible through online platforms or are proprietarily assembled through open-source code. Digital building blocks are transformational due to the innovative patterns that can be established through "use-recombination" [40], while there needs to be separate legitimacy for each building block, as well as collective legitimacy for a new institutional arrangement to emerge [21]. It may, for example, be that a platform-based building block holds legitimacy (e.g., a cloud-based AI facial recognition-system) because it performs within a predefined level of accuracy. However, for the organizational or wider institutional arrangement to gain legitimacy, the embeddedness of the building block into a socio-economic system needs to be accepted at a much broader level of implementation.

As digital building blocks are created by engineers, and as humans are subject to bias [45], this means that the values of the designer can be "frozen into the code, effectively institutionalizing those values" [37: 158]. Friedman and Nissenbaum [46] argue that bias in computer systems can arise in three distinct ways, referring to (1) pre-existing social values found in the ''social institutions, practices and attitudes'' from which a technology emerges, (2) technical constraints, and (3) emergent aspects that arise through usage, which only can be known ex-post. The distinction between social and technical bias has also been referred to as normative and epistemic concerns [47] or structural and functional risks [48]. Functional risks refer to technical areas such as the design and operation of an AI system, including datasets, bias, and performance issues, whereas structural risks refer to the ethical implications of an AI system, including the societal effects of automated decisions.

Based on a synthesis of the above considerations, I propose the use of three analytical constructs, referring to technological maturity, data, and AI-autonomy, in order to signify a field's relative elaboration of digital infrastructure. The constructs have been selected as they embody some of the main features of AI-induced digital infrastructure associated with (1) the algorithm, (2) its use of data, and (3) its ability to act, as well as the likely ramifications of those actions. Each of the three constructs are elaborated in greater detail below.

## 2.2 Technological Maturity, Data, and AI Autonomy

*2.2.1 Technological Maturity.* AI systems are subject to different degrees of maturity, both in terms of the accuracy of the system [49], as well as the elaboration of adjacent technological standards [50]. The accuracy of an AI-model refers to whether it operates within a predefined 'acceptable' level of performance. In the case of autonomous vehicle safety, for instance, an AI-controller is expected to hold the ability to locate persons and objects from a distance of 100 meters with an accuracy of +/- 20 cm, within a false negative rate of 1% and false-positive rate of 5% [51]. In some areas that involve high-stakes decisions (e.g., autonomous driving, credit applications, judicial decisions, and medical recommendations), high accuracy alone may not be sufficient, as these applications require greater levels of trust in their associated services [52]. In high-risk areas, it is important that the functional aspects of a model (i.e., accuracy, data, etc.) are further elaborated through measures such as certification, testing, auditing, as well as the elaboration of technological standards, which refers back to the institutional infrastructure of a field.

Depending on the context and the area of use, a range of quantitative measures can be used to evaluate the technological maturity of an AI-induced field. Some suggestions include the measures of scientific output, e.g., research papers, citations, and the intellectual property rights that surround a given field. Important questions relate to whether emerging algorithmic capabilities are under development and going through stages of testing or already are being widely deployed by a small or a large number of actors. For structural implications, it is important to ask questions such as: how does the technological maturity and elaboration (of immature/mature) AI-induced digital infrastructures affect a field? For example, the implementation of chatbots, which may have performed with sufficient accuracy under test environments, have proved to display racial biases and prejudices, as the algorithm continues to learn during actual implementation, which aggravates social harm for certain groups of the population [53]. The elements that are used to evaluate and decide whether an AI-system is mature or immature are therefore dependent on its context of implementation, which renders technical aspects alone insufficient when assessing the technological maturity of AI-models and associated digital infrastructure.

Several methods have been proposed to evaluate predictive models, such as "model cards for model reporting" [54], "nutrition labels for rankings" [55], "algorithmic impact assessment" forms [56], as well as "fact sheets" [52]. These frameworks can help organizations establish new organizational practices that characterize model-specifications in more coherent ways while paying special attention to attributes such as accuracy, bias, consistency, transparency, interpretability, and fairness, among others.

At a general level, when dominant standards are in place, and the accuracy of an AI-system is deemed safe, reliable, and trustworthy, digital infrastructure is considered elaborate, and higher field legitimacy is expected. If a technology is considered immature, inaccurate, or insufficiently tested, the surrounding digital infrastructure would be considered unelaborate.

*2.2.2 Data.* The nature of the data that feeds into any AI-model or system is also of particular importance, and data can be classified as being either sensitive (e.g., health-related) or non-sensitive (e.g., weather-related), and the nature of the data can be private (i.e., individual data) or public (common/pooled data) [57]. Data can also be biased, which makes AI systems prone to inherit either individually coded forms of bias or biases that result from historical or cultural practices, which are reflected in the training data, and could be adopted by the algorithm [58]. For an algorithm to be effective, its training data must be representative of the communities that it impacts. The use of digital infrastructures by individuals, machines, and communities, requires institutions to negotiate how bits containing varying kinds of information legitimately can be utilized and (re)arranged by organizations.

Several methods have been proposed to evaluate data as well as machine learning models under a variety of conditions. For data, these include "data statements" [59], "datasheets for data sets" [60], and "nutrition labels for data sets" [61], which seek to evaluate the data that goes into a model across training, testing, and post-implementation scenarios.

Sound data practices that are transparent, well-documented, and privacy-preserving, are generally associated with a more elaborate digital infrastructure. Data practices that are biased, undocumented, or otherwise disputed could be considered a sign of unelaborate digital infrastructure.

*2.2.3 AI Autonomy.* AI-agents hold varying degrees of autonomy to act, while the (explorative) actions of an autonomous learning agent may not always be known and can be subject to change depending on the data that is fed into the model [62]. An AI-agent can have limited or extensive autonomy to make decisions, while the decisions of an AI agent can have a lenient (e.g., recommender engine, smart speaker) or a severe (e.g., autonomous vehicle, incarceration system, facial recognition) impact on individuals as well as its surroundings, if the algorithm is inaccurate, fails, or is otherwise at fault. This could include aspects such as excessive collection of data or unwilling intrusion of privacy in the case of facial recognition systems, for example. The categorization of an agent's autonomy, therefore, includes its ability to act, as well as the possible ramifications of its actions. The perceived risk of an AI agent can be understood as the probability that a disruptive event occurs, multiplied by the severity of potential harm to an individual or form of organization [48]. The definition of "harm" and the computation of probability and severity is context-dependent and varies across sectors. For instance, the impact of an autonomous decision in medical diagnosis or in autonomous vehicles would, arguably, be greater than that of a product recommendation system [63]. Relevant questions include: what risks may be present in model usage, as well as identification of the potential recipients, likelihood, and magnitude of harms [62]. Where risks are taken into consideration and are sufficiently mitigated in relation to avoiding any potential harms, the digital infrastructure could be considered elaborate.

The elaboration of AI-associated digital infrastructure across the constructs of technological maturity, data, and AI-autonomy, remain subject to both qualitative and quantitative judgments and measures, which are field-dependent and linked to idiosyncrasies across functional (technical) as well as structural (ethical) risks and considerations.

## 2.3 Governance

Since field-level advancements in AI are context-dependent, this means that the existing institutional infrastructure and logics negotiates the actual impact that a technology is allowed to have within a given social context, which differs across geographies. In other words, the flexibility of a digital infrastructure is often restricted by socio-technical and regulatory arrangements (e.g., restrictions on autonomous vehicles, regulations on the use of patient's medical data, etc.). Oftentimes, layered and interoperable standards and common definitions of application and service interfaces guide the use and growth of digital infrastructures [64] and are necessary for digital infrastructures wider process of institutionalization. As large technology companies usually are the leading innovators of a field, these also carry a crucial weight in the direction of new technology standards [65], which generally affects how an industry or a field continues to evolve. Typically, private actors orchestrate ecosystems and associated digital infrastructures, which brings issues to the forefront, such as the challenge of establishing a governance system, reproducing social order, and incorporating aspects of value appropriation and control [64, 65, 48, 66, 67].

The process that renders digital infrastructures institutional occurs when innovators infuse specific norms, values, logics, as well as forms of governance and technological control into the infrastructure, and as the infrastructure becomes more widely adopted and legitimized over time [15, 68, 34]. Digital institutional infrastructure can thus be viewed as the integration of digital infrastructure and institutional infrastructure, which is defined as standard-setting digital technologies that enable, constrain and coordinate numerous actors' actions and interactions in ecosystems, fields, or industries [21].

## 3 A Conceptual Framework for Understanding AI-Induced Field Change

By integrating the insights from institutional theory (work, logics) with information systems theory (digital infrastructure), I propose the use of a novel framework for analyzing AI-induced field change (Table 1). The framework builds on Zietsma et al.'s [22] conceptualization of pathways of change, which hypothesizes how actors drive change across different sets of field circumstances. The proposed framework extends exisitng work [22] through incorporating the notion of AI-associated digital infrastructures, which has implications for the structure and organization of (digital) institutions going forward.

| ACTORS | |
|---|---|
| -Subject position: central, middle status, and peripheral actors -Characterized by roles or functions, i.e. field-structuring or governing organizations, formal governance units, field coordinators, etc. | |
| **DIGITAL INSTITUTIONAL INFRASTRUCTURE** | |
| -Standard-setting digital technologies that enable, constrain, and coordinate numerous actors' actions and interactions in ecosystems, fields, or industries [21]. | |
| **INSTITUTIONAL INFRASTRUCTURE** | **DIGITAL INFRASTRUCTURE** |
| Established through activities such as: certifying, assuring and reporting against principles, codes, rules, and standards, as well as through the formation of new associations and networks among organizations, including official rules and regulations [31]. **Logics**: refers to the relationships among individuals and organizations in the field. Logics can be competing or unitary. They may be based on market, social, and other considerations. **Work**: refers to the practices and actions of individuals and organizations that has implications for creating, maintaining, and disrupting institutions over time. Looks at the effect of institutional change on areas such as hierarchies of status and influence, as well as subsequent power relations. Incorporates the notion of field structuring events, which informs or disrupts logic formation. | Established from a multitude of digital building blocks, defined as the computing and network resources that allow multiple stakeholders to orchestrate their service and content needs [14]. **Technological Maturity**: refers to the elaboration of hardware and software-based infrastructures and associated technological standards. Includes the perceived accuracy, safety, and reliability of an AI system/agent. **Data**: refers to the data that is used in a model, which either can be sensitive or non-sensitive, private or publicly available, centralized or decentralized, and may be linked to varying forms of ownership. **Autonomy**: refers to whether the AI-agent holds limited or extensive autonomy to act, and whether the agent's actions have a negligible or a considerable impact on its environment and surroundings. |
| **GOVERNANCE** | |
| -Combinations of public and private, formal and informal systems that exercise control within a field. -Units and processes that ensure compliance with rules and facilitate 'the functioning and reproduction of the system (e.g., standards, regulations, and social control agents that monitor and enforce these). | |

**Table 1: Framework for Analyzing AI-Induced Field Change and Legitimization**

The framework first considers varying actors and their position in a field before elaborating on these ability to affect the direction of a field, either through the introduction of a new technology, regulation, or a social movement. Next, the relationship among actors as well as their coherency in terms of logics is considered. When logics are unitary, greater field alignment is expected, whereas competing logics means that a field is unsettled. The elaboration of institutional infrastructure is considered by looking at the practices and actions of individual actors as well as organizations in terms of creating, maintaining, and disrupting institutions over time. The notion of field structuring events is particularly important, both in terms of logic formation or disruption, as well as for the elaboration of the institutional infrastructure of a field.

The AI-associated digital infrastructure of a field is signified by the proposed constructs of technological maturity, data-specification, and the relative autonomy of an AI-system. Technological maturity refers to the perceived accuracy of an AI agent, as well as the elaboration of areas pertaining to standards, research, intellectual property, and so on. The data linked to a model is another important source of institutional legitimacy, both functionally (e.g., non-biased data) as well as structurally (e.g., how an organization is engaged in practices of data collection and usage). Autonomy refers to the relative impact of an AI agent on its general environment, as well as its potentials for exacerbating structural risks and create harm. At last, the governance of a field, as well as the mechanisms that guide algorithmic implementation, are considered.

Based on coherency in logics (unitary, competing) [19], and the elaboration of institutional infrastructure (high, low) [26], a four-fold classification of field conditions is produced around whether there are settled or unsettled logic prioritizations and limited or elaborated digital and institutional infrastructure (Figure 1) [22].

Where digital and institutional infrastructure is highly elaborate, and there is a unitary dominant logic within the field, the field can be described as established and relatively stable, i.e., the institutional infrastructure is coherent [22]. Formal governance and informal infrastructure elements are elaborate and likely to reinforce each other, leading to a coherent sense of what is legitimate or not within the organizational field [72, 73].

In fields where there is highly elaborate institutional infrastructure but competing logics (low coherency), there could be multiple formal governance and digital and institutional infrastructure arrangements [22]. These arrangements may be in conflict with one another or compete for dominance, which makes the field contested [25, 74]. Contested refers both to competing digital infrastructures (e.g., technological standards, varying models, and levels of algorithmic accuracy), as well as to stakeholders opposing views.

Fields with low coherency and limited elaboration of digital and institutional infrastructure are described as fragmented, with competing conceptions of what is legitimate. Fields may be fragmented if they emerge in intermediate positions (e.g., biotechnology), which draws on logics and practices from diverse but neighboring fields [74]. A field may also be fragmented as new actors enter an existing field with innovative ideas and designs about products, courses of action, behaviors, as well as new structures and ways of organization [75]. In the field of facial recognition technology, for instance, there are multiple competing logics that move across varying stakeholders and demonstrate incoherent views over technological accuracy, as well as the technology's inherent ability to enhance public safety. Many differing views paired with a limited (but expanding) digital infrastructure situates the field in the fragmented quadrant.

When infrastructure has a low degree of elaboration but a high degree of coherency in terms of unitary logics, the field is described as emerging or aligning [19]. While the lack of digital and institutional infrastructure in an emerging field may create considerable room for experimentation and change, it may also limit field members' ability to define and acquire legitimacy and thus contributes to ambiguity, and potentially, the need to draw on ill-suited infrastructure from adjacent fields. One example could be the emergence of autonomous vehicles, drawing on existing legal frameworks in terms of liability, which, however, are ill-suited in terms of covering the accompanying change in agency and responsibility.

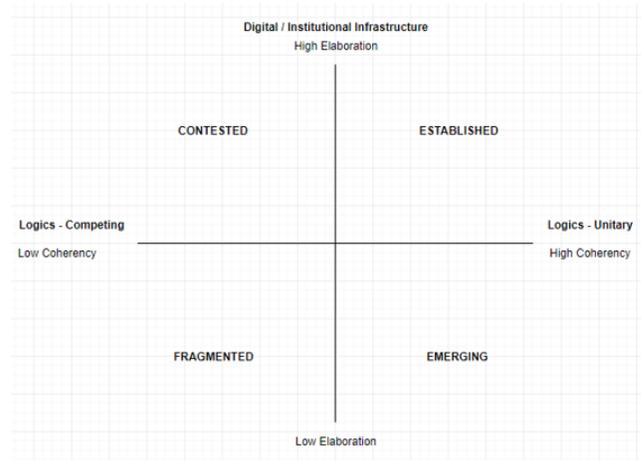

**Figure 1: Digital / Institutional Infrastructure & Logics: Framework for Field-Change (modified from [22])**

Categorizing a field's present condition as well as its potential trajectories enables us to get a deeper understanding of possible areas of contestation, fragmentation, or alignment, as well as what it takes for an AI-induced field to grow established over time. Before these conditions are further discussed in section 5, the following section applies the developed framework (Table 1) to the field of facial recognition technologies in the United States. The application briefly illustrates the utility of the framework in terms of assessing field-elaboration, while future studies may apply the framework to analyze specific case-studies at greater depths.

## 4 Analyzing AI-Induced Field Change and Legitimization: Facial Recognition Technology

### 4.1 Actors

The proliferation of facial recognition technologies in the United States has been supported by large technology companies, which are the central actors of the field (e.g., Apple, Amazon, Google, Microsoft, IBM). While these companies provide their own applications directly to the market, they also modularize facial-recognition technologies and make them accessible for complementors on their platforms. This makes them field structuring organizations since the modularization of FRT-systems embodies best-practices and de-facto industry standards, which other companies align with. Central actors include adopters of FRT-systems, while many of these are U.S public sector agencies. Contractors that specialize in delivering FRT-technology to law-enforcement agencies, as well as the National Institute of Standards and Technology (NIST), hold intermediate positions. Peripheral actors include multistakeholder organizations such as the Partnership on AI, non-profit research organizations such as the Center for Data and Society, as well as research institutes such as The AI Now Institute (NYU). These actors affect the field through public reports and commentaries, paying special attention to issues of technological implementation and social ramifications. Peripheral actors also include opponents of FRT-systems, both in the form of activists, as well as civil society organizations such as The American Civil Liberties Union (ACLU).

### 4.2 Logics

The dominant logics behind FRT's has been driven by private sector companies focused on gaining market share. The logics behind adoption is motivated by enhancing measures of public safety, e.g., in terms of identifying criminals, screening travelers, and processing border immigration. Both logics are highly contested by peripheral actors e.g., company activists and civil rights organizations [76], citing that inaccurate technologies hold the potential of exacerbating racial and social biases and inequities. This signifies that emergent dominant logics are at odds with existing social arrangements, including structures of power and governance, which makes the technology heavily resisted [77].

### 4.3 Work: Field Structuring Events

In 2019, the local government of San Francisco became the first city in the United States to ban the use of FRT's by local agencies. In the spring of 2020, nationwide protests against police brutality and racial profiling caused several central actors (IBM, Amazon, Microsoft) to stop providing FRT-technologies to law enforcement agencies altogether. IBM called for "a national dialogue on whether and how facial recognition technology should be deployed by domestic law enforcement agencies" [78: 1], and Amazon announced a one-year moratorium on police use of its facial recognition technology, giving policymakers time to set appropriate rules around the use of the technology. Microsoft declared that it would not sell FRT-technology to police departments in the United States until a federal law that regulates the technology is formulated. These actions by some of the central actors in the field signal that the existing institutional infrastructure remains inadequate in terms of governing and addressing the current expansion of FRT-related digital infrastructure. This indicates that even as central actors on the procurement side include many public sector agencies, the necessary institutional infrastructure to guide potential ramifications of immature technological adoption has not yet been formulated. Greater alignment between stakeholders across industry, government, and civil society, is currently needed in order to secure ongoing legitimacy as well as greater field-level elaboration and use of facial recognition technologies.

### 4.4 Technological Maturity

In terms of technological maturity, verification algorithms have achieved accuracy scores of up to 99.97% on standard assessments like the National Institute for Standards and Technology (NIST) Facial Recognition Vendor Test [79]. For identification-systems, error rates tend to climb when high-quality images are replaced with the feed of live cameras that normally are utilized in public spaces. Aging is another factor that affects error rates, while accuracies of FRT-systems differ considerably across gender and race [8]. The context, i.e., the specific area of implementation and use, can therefore be said to have wide-reaching consequences for the accuracy-rates of individual FRT-systems.

### 4.5 Data

Issues of legitimacy are also inherent in relation to the kinds of data that are being used for training FRT-algorithms. Many databases rely on publicly available face-annotated data, which in some cases are scraped directly from social media platforms and have raised issues over privacy and consent [80]. The company Clearview has, for example, assembled a database containing some 3 billion images, where many have been scraped from public-facing social media platforms [81]. This raises concerns about the legitimacy of data rights and usage, as well as the ability of existing institutional infrastructure to provide, and safeguard, associated rights. The quantity of data is in many cases important for algorithmic training, as well as for retaining levels of accuracy post-deployment, which means that there is an inherent incentive for private, as well as for public-sector adopters, to amass rich databases (e.g., new biometric data), in order to increase and continue to ensure the accuracy of a given system. In several states (e.g., Texas, Florida, Illinois), the FBI is allowed to use facial recognition technology to scan through the Department of Motor Vehicles (DMV) database of drivers' license photos [82] in order to generate a more coherent centralized biometric database. As these kinds of data contain personal information, they are classified as being sensitive and vulnerable, both in terms of misuse as well as in relation to cybersecurity breaches and possible identity theft [57].

### 4.6 Autonomy

AI in facial recognition-systems is perceived as a new kind of social control agent, which may exert autonomy over law-enforcement officers in relation to issuing arrest orders. If the accuracy of a system is flawed, an officers' actions are likely to cause social harm whenever an innocent citizen is arrested [83]. The adoption of facial recognition systems for use in law enforcement alters existing power dependencies, as officers have to trust in, and act on, the information that is rendered to them by the system. Facial recognition systems are thus shaping entirely new practices and forms of organization in which the autonomy of the AI-agent is dependent on the delivery of accurate information, which could reinforce a drive towards data-centralization.

### 4.7 Governance

The field of facial recognition technology is fragmented and exhibits low coherency and limited elaboration in terms of institutional infrastructure. A lack of governance is most readily seen in the absence of coherent rules and regulations, while the field is currently going through a shift from self-regulation towards more formalized governance arrangements. This shift has been called for by peripheral actors, and more recently also by central actors from the private sector, which demands new rules to guide legitimate implementation going forward. The case of facial recognition technologies used by law-enforcement highlights the critical role of culture and politics involved in the organization of markets and in creating the governing 'rules of the game' [84, 85, 86].

## 5 Pathways of Change: How AI-Fields Move and Gain Legitimacy

In order to move from a static to a more dynamic analysis of the conditions related to field change, this section applies the concept of 'pathways of change' to a number of distinct areas of AI. As evident, each area of AI implementation is subject to idiosyncrasies that are linked to a field's specific form of digital and institutional infrastructure, as well as their elaboration. Pathways of change suggest that there are some commonalities to how fields are likely to evolve and where obstacles to legitimization and institutionalization may be found. In order to understand how fields move between states, special attention needs to be placed on the scope of change (i.e., which elements change and how much changes)[87], as well as the pace of change (i.e., the speed at which a field moves from one condition to another)[88].

In the case of facial recognition technologies, the field is currently moving from the fragmented towards the contested quadrant, as the number of use-cases (e.g., public surveillance, airport check-ins, smartphones, doorbells, etc.) continues to expand, based on rising technological maturity (e.g., accuracy, standards). While digital infrastructures are expanding, the field continues to be represented by incoherent logics and sparse institutional infrastructure, however. For example, verification-based FRT's (e.g., unlocking a smartphone) is already a well-established practice that exhibits legitimate institutionalized functions. Identification-based FRT's (e.g., public surveillance), on the other hand, are more likely to stay contested due to having a lower degree of algorithmic accuracy, which is paired with more severe social impacts linked to the autonomy of AI-agents, and how these alter existing power structures. In order for the field, as a whole, to grow more established, a shift from self-regulation towards formalized governance arrangements and greater alignment and coherency in terms of logics is needed. In more authoritarian settings, such as in China, the field of facial recognition is already on its way to becoming established. This signifies that a country's socio-political setting informs its institutional infrastructure, which has important implications for a technology's path towards legitimization as well as processes of institutionalization.

A pathway that moves from an aligning or emerging field condition to an established condition usually involves a process of convergence, which is commonly observed in the institutionalization of most fields (see, e.g., [89]). The field of autonomous vehicles (AV), for example, is characterized by its emerging digital and institutional infrastructure, which has a low degree of elaboration but some coherency in terms of logics. While the field is currently aligning at a relatively slow pace, it develops in extension of an existing field (auto-infrastructure) that has been elaborated over decades. Large parts of the existing infrastructure are challenged, however, through the introduction of novel AI-agents and a transfer in autonomy from humans to machines. As the digital infrastructure is further elaborated, which entails a greater number of mixed-autonomy vehicles on the road, the field could move towards the contested quadrant, as logics associated with safety and liability are disputed. If the rules and regulations to handle negative externalities brought about by algorithmic errors are not in place, the field would likely stay in the contested quadrant. As the advent of AV's is going to shift the terms of liability [90], the scope of change demands that an entirely new institutional infrastructure has to be developed and elaborated by insurers, policymakers, legislators, and automakers, which could take years and be subject to multiple areas of contestation among stakeholders.

Another common pathway is the movement from an established to a contested field condition. This move is likely to occur through more disruptive change, either an exogenous shock, e.g., new regulation, or a strong social movement, or through the challenging of status quo by a new or peripheral actor [94, 95]. The use of recommender engines (RE), which suggests products, services, and other online information to users based on prior data, is already a well-established practice but could grow more contested due to incoherencies in logics. RE's have, for example, been argued to create fragmentation by limiting a users' media exposure to a set of predefined interests or objectives [93], which could have undesirable societal consequences as people's preferences may be guided towards echo chambers, where alternate views are missing [94], which further impedes decisional autonomy [95]. Other actors argue that existing data are inconclusive, and some research suggests that recommenders appear to create commonality, not fragmentation [96], implying that there is little cause to modify the current architecture of recommender engines [97, 100]. This incoherency in logics is coupled with information asymmetries between the AI-agent and human actors in relation to how, and on which information, a decision to recommend certain content is rendered. This lack of transparency, as well as a lack of algorithmic knowledge by the general population, arguably leaves certain elements of the current digital infrastructure in the contested quadrant. The governance of data and information that goes into a recommender engine, for example, is partially situated in the contested quadrant, which could have wider field-level implications, and possibly force a coercive change in the form of new regulation [98].

When a field moves from a position of established to (re)aligning under the emergent quadrant, change is usually observed through incremental modifications, with central actors often managing these [22]. This incremental change sees the field realigning around new practices or relational channels while readjusting the institutional infrastructure. The field of smart speakers (Google Assistant, Siri, Alexa, etc.) has moved from the emerging towards the established field-quadrant over a relatively short time-horizon, while certain elements of the digital infrastructure have been linked to concerns over data-collection and data privacy practices, which could see the field move to grow more contested.

Other pathways of change include a move from a fragmented or contested condition to one that is aligning in the emergent quadrant. When looking at nascent AI areas such as Generative Pre-trained Transformer 3 (GPT-3), or deepfakes, these fields emerge in the fragmented quadrant due to incoherent logics, coupled with institutional infrastructures that are unelaborate. While the inherent agency of these AI systems are emerging, their associated use of already elaborate digital infrastructure linked to the general information ecosystem makes them able to proliferate at rapid speeds. In terms of autonomy, this means that these AI-agents could have a considerable impact on their environment by exacerbating the spread of misinformation online. A move from the fragmented quadrant towards greater alignment is therefore needed, which may be formed as actors converge around new ideas, rules, and positions in order to inform and elaborate the surrounding institutional infrastructure while establishing greater coherency in logics [89, 102].

AI is currently changing organizational practices across a wide range of fields, which implies that new applications should be carefully considered in terms of their short-term impact on human behavior as well as long-run influences on institutional change. Insufficiently tested implementation of unsafe or biased algorithms can foster negative externalities, which can have severe consequences or may be detrimental to societal trust. An analysis of AI-associated digital institutional infrastructure, based on logics and work, as well as conceptualizations of technological maturity, data, and AI-autonomy, contributes to assessing where potential areas of contestation or fragmentation could be found. These findings hold important implications for AI-developers and adopters (e.g., engineers, managers, firms), as well as for policymakers that seek to define new rules going forward. These implications, as well as the main takeaways of the paper, are briefly discussed below before a conclusion is offered.

## 6 Discussion: Commonalities of AI-Induced Field Change & Pending Issues over Governance

Through illustration of the developed framework, three takeaways which move across varying kinds of AI-induced field change and legitimization, are offered. Subject to discussion, these broadly refer to (1) altered power-dependencies between humans and machines, (2) unresolved questions over data-use and control, as well as (3) issues with the current elaboration of institutional infrastructure surrounding many forms of AI application.

First, the autonomy of AI agents can affect existing power-dependencies, which may cause friction as human behavior and ways of organization are influenced in ways that are hard to identify ex-ante [35]. In examples such as facial recognition, judicial AI-systems, autonomous vehicles, and so on, the AI-agent gains determining power over human actors, which have to trust the identifications or predictions of the AI-agent. This transfer of autonomy is contingent on systemic trust, which is based on conceptualizations of technological maturity and ideas of machine-augmented perception that is expected to operate at cognitive levels that are equal to – or in many cases exceeds those of a human operator. Issues with field-level legitimization and nascent processes of institutionalization are therefore likely to arise when emerging systems are inaccurate, unsafe, or intransparent, which erode trust across applications and causes fields' to stay fragmented and logics to grow incoherent. Analyzing the field trajectories of such cases, involves assessing what it takes for altered power-dependencies to be conceived as legitimate practices, which is crucial for a field to move from fragmentation or contestation towards greater alignment of digital and institutional infrastructures.

Second, an incentive for data-centralization is inherent in most digital infrastructures (based on technical and economic logics), which has implications for associated forms of organization. A lack of transparency during the processes of data collection, as well as in markets for data, are leaving large populations unaware of where and how their personal data and information is being used, stored, and traded, as well as for what purposes [47]. The current organization of many digital infrastructures thus come with the risk of deteriorating public trust in digital institutional infrastructures if data-sources are used for socially disputed measures of public (e.g., safety) and private (e.g., market-based) forms of surveillance [12], or are being misused, e.g., due to large-scale data-breaches [101]. This implies that the legitimacy of AI-agents is highly contingent on legitimate collection, use, and ownership of data, which otherwise could be a source of dispute that causes field-level disintegration. Regulations such as the European Union's General Data Protection Regulation (GDPR) should be seen as the first step of elaborating institutional infrastructure that seeks to move fields engaged in data-collection from the contested quadrants towards greater establishment and coherency in logics. Over time this could imply a conceptual shift of companies moving from "owners" towards "custodians" of individuals' private data. Opening access to data and developing interactivity, as well as an increased sense of ownership with users, is a step that could gain traction in order to smoothen existing information asymmetries between central actors and individual end-users [102]. Similarly, empowering users to better understand and perhaps interact with certain AI-agents (e.g., recommender engines) would empower these with a greater sense of ownership over how streams of information are utilized and handled, as well as impacting individual practices and ways of behavior.

Third, where institutional infrastructure is considered inadequate during phases of market expansion, peripheral actors, such as civil society organizations, frequently work on outlining insufficient governance arrangements [103]. In many cases, it is important that institutional infrastructure is elaborated before negative externalities start to erode systemic and institutional levels of trust, which causes a field to grow fragmented. If trust is eroded past certain barriers, technology developers and adopters are likely to experience severe pushback from the general public.

Public pushback forces central actors from the private sector to engage in new measures of self-regulation, which in some cases means scaling back digital infrastructure until a policy-vacuum is filled by new legislative provisions, such as in the case of facial recognition technologies in the United States. When logics are at odds with existing power structures or violate existing governance arrangements, these are also more likely to be resisted [77].

At the same time, the formulation of institutional infrastructure needs to emerge in more adaptive forms of organization [104, 105], that are able to take into account the myriad ways in which modular AI-systems influence and shape existing practices and ways of behavior. This warrants that new types of institutional engineering have to be embraced in order to keep up with rapidly expanding digital infrastructures while alleviating the pacing problem [7]. Proposed measures of institutional adaptation to mitigate AI-induced externalities include enhanced measures of algorithmic auditing carried out by companies [106], third-party auditors [107], or external regulators [108].

Auditing can create an ex-post procedural record of complex algorithmic decision-making in order to track inaccurate decisions or to detect forms of discrimination, as well as biased data, practices, and other harms [47]. When algorithms are designed without considering a population's or community's needs, it has become apparent that both the algorithm and its implementer are likely to experience public pushback or outright rejection, which may obstruct other processes of AI legitimacy and adoption [109].

As a growing number of fields continue to migrate from traditional forms of linear programming and further embrace autonomous learning algorithms – behavioral control is gradually transferred from the programmer to the algorithm and its operating environment [110]. During this process, "the modular design of systems can mean that no single person or group can fully grasp the manner in which the system will interact or respond to a complex flow of new inputs" [111: 14]. In order to cope with AI-induced complexities, new governance structures have to be co-invented through greater stakeholder engagement among companies, civil society organizations, as well as policymakers in order to secure the inclusion of affected communities in the development of just algorithmic systems and processes going forward [112].

The tradeoffs between algorithmic accuracy, transparency, and use of data, as well as the rights to privacy, explanation, and right of redress, remain subject to ongoing forms of mediation in relation to the concomitant organizational practices that emerge at the intersection of human-machine-based interactions. While these tradeoffs have wide-ranging implications for the kind of institutions that are likely to emerge, the devising of inclusive yet reflexive institutional infrastructures that are able to encompass a wide variety of AI-associated risks remains a crucial area to be studied for years to come.

## 7   Conclusion

The increased presence of AI-agents embedded in varying forms of organization entails that a whole range of AI-induced institutions are currently emerging. This paper makes three contributions that help elicit the ways in which AI-induced fields are subject to varying degrees of legitimacy as well as processes of institutionalization. First, the paper proposes a novel conceptual framework for analyzing AI-induced field change. Second, it illustrates the utility of the framework and finds a set of common grounds for contestation associated with AI-induced field change and legitimization. Third, the paper points to the need for more adaptive organizations to emerge in response to the rapidly evolving digital infrastructures of AI-systems.

The notion of pathways of change helps elicit the varying ways in which novel AI solutions are resisted, rejected, or accepted as legitimate practices over time. Assessing where a field is currently positioned, as well as what its potential trajectories are, or could be, and what needs to be done for a field to grow established and become legitimatized over time, are essential considerations for stakeholders to take into account. Such deliberations contribute to secure greater alignment between digital and institutional infrastructures, which is important in terms of mitigating negative externalities going forward.

The logics of any algorithmic interaction, as well as transparency with the information that guides the interaction, needs to be broadly examined in order to get a better understanding of how a given AI-agent affects and potentially alters existing dependencies between humans and machines, as well as between humans and new forms of organization. Only by understanding where certain negative externalities could potentially arise can organizations that are responsible for algorithmic development or implementation work on establishing the necessary institutional infrastructure (i.e., standards, rules, and processes) to keep such externalities in check. Transparent and reliable AI systems, as well as enhanced human-AI interactions, is a key element for the trajectory of most AI fields on their road to secure a broad sense of social legitimacy as well as growing established over time. As novel digital infrastructures continue to emerge, it is important that their road to becoming institutionalized structures of society is thoroughly vetted and mitigated in order to secure fair, equitable, and trustworthy socio-technical interactions in the years to come.